\documentclass[a4paper]{article}

\usepackage[a4paper,margin=3cm]{geometry}
\usepackage{graphicx}
\usepackage[hidelinks,bookmarks=false]{hyperref}
\usepackage{multirow}
\usepackage{enumitem}
\usepackage{tabularray}
\usepackage{authblk}
\usepackage[square,numbers]{natbib}
\usepackage{amsmath}

\title{Fast Multi-Party Open-Ended Conversation\\ with a Social Robot}

\author{Giulio Antonio Abbo\footnote{These authors contributed equally.}}
\author{Maria Jose Pinto-Bernal\protect\footnotemark[1]}
\author{Martijn Catrycke}
\author{Tony Belpaeme}
\affil{\textit{IDLab-AIRO}, \textit{Ghent University -- imec}, Ghent, Belgium}
\affil{\texttt{giulioantonio.abbo@ugent.be, mariajose.pintobernal@ugent.be}}

\date{\scriptsize January 2025 -- Under review}

\begin{document}

\maketitle

\begin{abstract}
Multi-party open-ended conversation remains a major challenge in human-robot interaction, particularly when robots must recognise speakers, allocate turns, and respond coherently under overlapping or rapidly shifting dialogue. This paper presents a multi-party conversational system that combines multimodal perception (voice direction of arrival, speaker diarisation, face recognition) with a large language model for response generation. Implemented on the Furhat robot, the system was evaluated with $30$ participants across two scenarios: (i) parallel, separate conversations and (ii) shared group discussion. Results show that the system maintains coherent and engaging conversations, achieving high addressee accuracy in parallel settings ($92.6\%$) and strong face recognition reliability ($80\text{--}94\%$). Participants reported clear social presence and positive engagement, although technical barriers such as audio-based speaker recognition errors and response latency affected the fluidity of group interactions. The results highlight both the promise and limitations of LLM-based multi-party interaction and outline concrete directions for improving multimodal cue integration and responsiveness in future social robots.
\end{abstract}

\section{Introduction}

Rapid advances in both hardware and conversational AI have accelerated progress towards social robots capable of engaging in natural, fluid conversation with humans in a natural and intuitive way~\citep{obaigbena2024ai,schobel2024charting}. As a result, robots are increasingly deployed in education~\citep{belpaeme2018social,verhelst2024adaptive}, healthcare and therapy~\citep{kyrarini2021survey,cifuentes2020social}, as well as for companion roles ~\citep{broadbent2023enhancing,giudici2024delivering}, where expectations for rich social and conversational behaviour continue to grow. At the same time, embodied conversational agents have achieved notable improvements in gesture generation, expressiveness, and user engagement~\citep{wolfert2022review, aneja2021understanding, gibson2000seizing}. However, this progress has not translated equally across interaction types. While dyadic human-robot conversation has seen substantial advancements, open-ended multi-party interaction, defined as spoken interaction involving more than two participants, remains a challenge, especially when the robot must coordinate naturally with several users at once.

Multi-party interaction introduces complexities that go far beyond a dyadic conversation. Robots must identify who is speaking, distinguish overlapping utterances, determine the intended addressee, and manage turn-taking within rapidly shifting conversational dynamics~\citep{addlesee2024multi,foster2012two,moujahid2022multi,murali2023improving,paetzel2023improving,richter2016you,zarkowski2019multi}. These requirements challenge perception, timing, dialogue management, and real-time adaptation.  Even with recent advances in large language models (LLMs), most existing systems manage these situations by constraining turn-taking or limiting user behaviour, which restricts the naturalness and realism of the interaction.

While recent progress in LLMs has greatly strengthened the conversational abilities of social robots, enabling more flexible, context-aware, and human-like dialogue~\citep{chang2024survey}, these advances have been demonstrated primarily in controlled, single-speaker settings. Multi-party conversation remains mostly absent from LLM research and deployment, leaving open the question of whether these models can reliably track multiple interlocutors, maintain distinct conversational threads, or adapt to disruptions such as rapid turn switches and fragmented discourse.

Our work contributes towards this gap by exploring how multimodal perception and LLM-driven dialogue generation can support multi-party interaction. We present a conversational architecture that integrates voice direction arrival, speaker diarisation, and face tracking with an LLM-based dialogue manager to enable open-ended exchanges between a robot and two users interacting at the same time. The system was implemented on the Furhat social robot~\citep{al2012furhat}, leveraging its expressive face projection and microphone array for multimodal perception. 

To evaluate the system under different conversational dynamics, we conducted a controlled multi-user evaluation in which two participants interacted freely with the robot in an open-ended conversational setting. This evaluation design allowed us to examine the capabilities of the system in a realistic multi-user setting. Building towards this goal, our study addresses the following research questions.
\begin{itemize}
    \item \textbf{RQ1} How effectively can the system maintain multi-party conversations, particularly in terms of managing turn-taking and recognising participants under different conversational dynamics?
    \item \textbf{RQ2} What technical barriers, such as latency or recognition errors, most affect interaction quality, and how do these influence user engagement?
    \item \textbf{RQ3} What aspects of the system's performance do users identify as needing improvement, and how do these insights align with our quantitative metrics?
\end{itemize}

\section{Background and Related Work}

Managing effective conversations involving more than two speakers has long been an important goal.
In this section, we present an overview of the main challenges in multi-party conversations, drawing on our experiences and previous work in the field, and highlight the importance of certain cues in conversational agents.
Then, we analyse how three recent related studies approached multi-party conversations.

\subsection{Challenges}

\citet{addlesee2024multi} recently approached the problem of multi-party conversations.
In their work, they mention several challenges that increase the difficulty of this task, namely speaker recognition, turn-taking, group tracking, and goal tracking.

\textbf{Speaker recognition} is the task responsible for identifying the person speaking.
This increases the accuracy of the transcribed dialogue ---as dialogue is attributed to the correct speaker--- which in turn improves the relevance of the generated responses.
Speaker recognition can be achieved through voice or face recognition.
Through voice recognition, the speech is analysed, and the speaker is identified in a relative or absolute way~\citep{park2022review}.
Relative voice recognition labels each recorded utterance with a user ID valid within the conversation.
However, one person might receive multiple IDs throughout the conversation.
Absolute speaker recognition solves this issue by uniquely identifying users across conversations through a prerecorded sample of their voice.
Face recognition enables absolute speaker recognition~\citep{adjabi2020past} by extracting features from the face and building a user profile, to which detected faces can be compared in search of a match.

\textbf{Turn-taking} is the second challenge.
People perform it effortlessly, while for a conversational agent this task is still challenging~\citep{skantze2021turn}.
A conversational agent should find occasions to contribute to the conversation in a non-intrusive manner.
However, in practice, identifying such locations in the conversation relies on the analysis of many aspects, such as eye gaze, pragmatically complete sentences, prosody, or even breathing.
Relying on pauses between sentences has proven to be an unreliable method~\citep{skantze2021turn}, as in-turn pauses are often longer than their inter-turn counterparts.
Indeed, many turn switches occur through terminal overlap and interruptions, which should be handled appropriately.
Finally, next-speaker selection is an important aspect of turn-taking that must be considered.
Determining the addressee of a spoken utterance gives insight into whether the robot should respond, and is fundamental for following the conversation and contributing effectively to it.

The challenge of \textbf{group tracking} involves investigating the group dynamics.
For instance, detecting whether the group of conversational participants consists of an actual group or separate individuals having different conversations -- a situation not uncommon at a busy help desk, to give an example.
In this work, we investigate whether an LLM can handle both cases.

The last challenge mentioned is \textbf{goal tracking}.
To effectively perform goal tracking, the participants' goals are determined throughout the interaction to improve the generated responses.
This becomes difficult in multi-party conversations as people might have shared goals, different goals, or talk about other people's goals~\citep{murali2023improving}.

However, there are many more challenges in handling group interactions.
For instance, in their scoping review of computational challenges in social group human-robot interactions, \citet{nigro2024social} highlight \textbf{engagement detection} as another important topic.
Indeed, engaging each conversational participant equally is important to facilitate an effective and successful conversation~\citep{murali2023improving,moriya2012estimation}.
According to their findings, the most used features to measure engagement are audio features, such as prosody, and video features, such as gaze direction.
Furthermore, they report how disengagement often occurs in groups where participants exclude the robot and only continue interacting with each other.
Including anthropomorphic characteristics in robots is related to improved engagement~\citep{mehmood2024embracing,ghiglino2021mind}. In this both verbal and non-verbal aspects contribute to this perception~\citep{chidambaram2012designing}.

The \textbf{verbal component} of improved engagement requires a complete understanding of the interaction and the ability to generate human-like responses~\citep{gibson2000seizing}.
These responses should be natural in their structure and sound in their contents.
Furthermore, the latency with which responses are generated must be considered, as long waiting times contribute negatively to the perceived agency~\citep{skantze2021turn}.

At the same time, the \textbf{design and behaviour} of the embodied agent will influence the perceived agency.
Aspects of turn-taking behaviour, breathing, or other normally unconscious behaviours contribute to this perception.
Consequently, it is important to model the robot's non-verbal aspects consistently~\citep{chidambaram2012designing}.
Inconsistencies and mismatches between these aspects lead to what is called \emph{uncanny valley}~\citep{mori2012uncanny}, which entails not only lower perceived agency but often a sense of revulsion.

\subsection{Related Work}

Before the advent of LLMs, approaches to multi-party dialogue were grounded in theories developed primarily for two-party interactions, which were then adapted and extended to account for the additional complexity of group settings~\citep{branigan2006perspectives}.
Researchers like \citet{clark1992dealing} proposed that multi-party dialogues are governed by the same foundational principles as two-party ones, though the dynamics are more complex due to the presence of side participants and multiple potential addressees.
Among these principles, the collaborative accumulation of common ground and the Principle of Responsibility play an important role.
These state that each conversation participant is responsible for keeping track of the conversation contents and for allowing other participants to do the same, while at the same time speakers collaborate with addressees directly, and indirectly with side participants to achieve mutual understanding.
In practical systems, this theoretical distinction was operationalised by assigning and tracking participant roles~\citep{clark1982hearers}: speaker, addressee, and side-participant.
Using explicit heuristics, dialogue tags, or interaction context, alongside simplified interactions and strong dialogue policies, allowed the system to model obligations and expectations for grounding and turn-taking.

The inherent common ground and contextual understanding of LLMs overcome many of the technological limitations of previous implementations.
For instance, \citet{addlesee-etal-2023-multi} investigated how LLMs can be used to perform multi-party goal tracking in hospital settings.
They introduced a corpus of $29$ multi-party conversations between patients, their companions, and a social robot in a memory clinic.
The conversations are annotated to track the goals of the participants.
They evaluated different models and prompt techniques.
Among these, GPT-3.5-turbo, when guided with reasoning-style prompts, significantly outperformed the other models.
This kind of goal tracking is fundamental in task-oriented multi-party conversation, while, for conversational-oriented systems, dialogue flow or speaker recognition might be enough to fill in the gaps and provide context-aware robot behaviour.

In a later work on a similar multi-party setting, \citet{addlesee2024multi} achieved satisfactory results using the open-source LLM Vicuna for response generation and turn-taking features.
Their goal was to provide a system that is practical and entertaining within the context of healthcare clinics.
As this LLM runs locally, it avoids network traffic latencies and privacy concerns.
The authors, however, did not implement non-verbal features of the robot, which reduces the robot's capacity to achieve realistic turn-taking during the interaction.
They also did not implement speaker recognition, which makes it harder for the conversational AI to produce useful responses toward the specific speaker and track this person throughout the interaction.

\citet{murali2023improving} created a meeting facilitation robot for group decision-making.
Although this system was not created to handle open-ended conversations, it can handle other aspects of multi-party conversations, such as diarisation and participant engagement.
Their findings showed that confidence levels increased in groups that had the robot mediator.
The effectiveness also increased as the robot ensured that participants stayed on target.
However, the robot did not actively participate in the conversation, reducing the need for accurate turn-taking capabilities.

The evolution of multi-party dialogue systems reflects a shift from rule-based, heuristic-driven approaches to the more flexible, context-aware capabilities offered by modern LLMs.
While early systems relied on explicit role assignment and simplified interaction models, recent advancements leverage the inherent contextual understanding of LLMs to address challenges like goal tracking, speaker recognition, and engagement detection.
However, gaps remain, particularly in integrating non-verbal cues, managing real-time turn-taking, and achieving fluid conversations.
These limitations underscore the need for holistic systems that combine the strengths of LLMs with robust perceptual and behavioural modelling.

\section{Proposed Multi-Party Interaction System}

\subsection{Requirements}

The design and implementation of the multi-party interaction system was guided by the goal of achieving open-ended, multi-party spoken interactions, from which several requirements follow.

The implementation must track the ongoing discourse and match each participant with their contribution.
The first requirement is necessary for handling interactions without a specific goal, where the exact words spoken can influence future conversation turns.
The second follows from the goal of interacting with multiple speakers, including the possibility of addressing them individually, which requires accurately identifying the author of each statement.

The system must be capable of contributing to the conversation, either through pertinent comments or relevant questions on the topic discussed.
To achieve this, the system must understand when it is most appropriate to intervene and when to yield the turn to other participants, while also handling interruptions and barge-ins.

Additionally, the system must display non-verbal cues, such as gaze and eye movement, which improve the handoff to another speaker.

Finally, the implementation must meet three additional requirements.
First, minimising delays is critical to ensure responsiveness and maintain the flow of natural conversation.
Second, the system must be designed for expandability to accommodate future advancements or additional functionalities.
A modular architecture allows for the integration of new capabilities without requiring a complete redesign, ensuring long-term adaptability of our system beyond the scope of its evaluation.
Third, the system should remain independent of the embodiment's hardware to ensure versatility across different platforms and physical forms, as we do not want to be constrained by specific hardware limitations.

\subsection{Components Overview}

To satisfy the requirements previously introduced, we implemented a modular, event-driven software architecture that allows easy integration of new features and flexibility across different hardware platforms.
Each module in the system is designed to handle a specific set of tasks associated with multi-party interactions, such as speaker awareness, verbal and non-verbal interactions, voice recognition, face tracking, speaker diarisation, and turn-taking.
A conversation manager module integrates all of these elements to orchestrate the system's behaviour. A description of each module is provided below, while additional implementation details are reported in the Supplementary Material.
Figure~\ref{fig:flow} shows an example of the information flow between components.

\begin{figure}[!ht]
    \begin{center}
        \includegraphics[width=\linewidth]{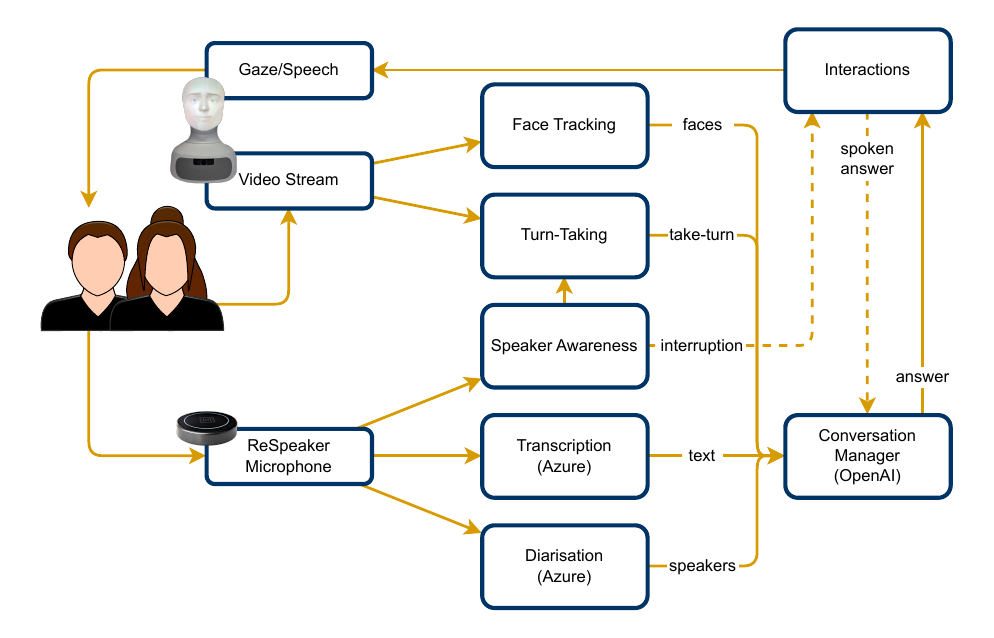}
    \end{center}
    \caption{Simplified overview of the information flow between components. The Conversation Manager receives information on the faces and speakers detected, the transcribed text, and whether it should take the turn. Then, it produces an answer and an addressee that are returned to the robot. In case of a user interrupting the robot (dashed lines), the system saves in the conversation only the part of the answer that was actually spoken.}
    \label{fig:flow}
\end{figure}

The \textbf{Speaker Awareness Module} uses an external microphone to determine the direction of the speaker's voice, thereby identifying when conversational turns shift among participants and the robot. A ReSpeaker\footnote{\url{https://wiki.seeedstudio.com/ReSpeaker-USB-Mic-Array/}. We observed that using the manufacturer's microphone case makes its direction of arrival data completely inaccurate, and as such, removed the casing for our study.} microphone array is used to perform this task and enhance the audio capture, which helps the system to accurately identify speakers during multi-party conversations. This module generates events such as detected speaker angles, turn changes, and the start or end of a user or robot's speech. 

The \textbf{Transcription Module} transcribes spoken words in real time using Azure Cognitive Services' speech-to-text capabilities. This module continuously records and transcribes each utterance and keeps track of the last spoken word to manage the conversational flow. By integrating with the speaker awareness module, the system avoids picking up and transcribing its own speech. 

The \textbf{Interactions Module} manages both verbal and non-verbal interactions. It leverages the robot's -- in our case, the Furhat robot -- text-to-speech capabilities for verbal responses. For non-verbal cues, it controls the eye gaze and head movements, looking towards the location of the user currently talking. This module can also handle interruptions, pausing the robot's speech when users begin talking. Although Furhat can display facial expressions, this study focuses primarily on gaze and head movements as interaction tools.

The \textbf{Diarisation} and \textbf{Face Tracking Modules} are responsible for identifying participants through voice and face recognition. Voice recognition uses Azure Cognitive Services to uniquely identify users' voices across interactions, while face tracking utilises Furhat's built-in capabilities along with Python's face recognition library\footnote{\url{https://github.com/ageitgey/face_recognition}} to recognise users across multiple sessions. This enables continuity in user interactions and allows for personalisation in future engagements, such as user preferences and history of previous interactions.
Azure's services require an initial enrolment phase for each new conversation participant. This phase consists of reading a sentence out loud so that the system can obtain a sample of the user's voice. The sentence is the same for all users, and the enrolment happens before the actual interaction begins. A similar process is used to enrol faces.
The system combines voice and face recognition based on the voice direction of arrival and the face position in the video frame. In the case of conflicts, priority is given to face recognition as it has proven to be more reliable.

The \textbf{Turn-Taking Module} uses data from the other modules, including face orientation and user position, to determine when the robot should take or yield a turn in the conversation. Face orientation cues play a critical role in the system's decision-making process regarding turn-taking. The system takes the turn if the robot is looked at by the last speaker or after a prolonged silence.
The use of silence instead of more sophisticated techniques~\citep{skantze-turn-taking} is justified with the assumption that the users are going to pass the turn through gaze. We plan to experiment with these techniques in the future. In case of failure, when a user starts talking while the robot has initiated its turn, the Interactions module pauses the robot's speech and gives the turn to the user. If the user stops as well, then the robot resumes from where it was stopped.

The \textbf{Conversation Manager Module} acts as the central control point, integrating the outputs of all other modules. Using GPT-3.5 (the latest version available at the time of this study) for response generation, this module ensures timely and contextually relevant contributions, managing the overall flow and content of the conversation.
We used the \emph{stream} option of OpenAI APIs, meaning that the response is immediately available as soon as it is generated.
Currently, to generate an answer, the system uses only the transcribed text; in the future, we plan to extend support to images from the camera stream for improved contextual awareness~\citep{10973830}.

The system is deployed on the Furhat robot~\citep{al2012furhat}, a talking head with advanced facial expression capabilities projected onto the robot's face. Furhat is equipped with a built-in camera for detecting and tracking faces, high-quality speakers for auditory output, and text-to-speech capabilities that allow it to generate expressive and natural-sounding speech. Additionally, Furhat has a customisable face that can be projected to convey different personas and emotions, making it ideal for engaging in social interactions, which is crucial for studying multi-party interactions. The robot allows for flexible control of gaze and head movements, which ensures effective non-verbal communication cues. These features make Furhat a suitable platform to evaluate our system.


\section{Methodology}

\subsection{Procedure}

This study employs a multi-phase design to evaluate the social and conversational performance of the proposed multi-party interaction system deployed on the Furhat robot. The evaluation was designed to capture both technical performance and user perceptions across two distinct interaction settings.

Participants were recruited in pairs to ensure familiarity and promote natural, fluent conversations. Prior to the experiment, participants provided informed consent in accordance with the guidelines approved by Ghent University's Ethics Committee. A pre-test questionnaire was used to assess participants' prior perceptions and knowledge of social robots. The pre-test included items on robot familiarity, general attitude toward technology, and prior exposure to AI systems.

A short enrolment step followed, required for Azure's identification service. Each participant read a short scripted sentence aloud, enabling the system to create a voice profile. Participants were then instructed to speak naturally, address the robot freely, and interact with each other as they normally would. No restrictions were placed on interruptions, speaking order, or conversational flow to preserve ecological validity.

\begin{figure}[!ht]
    \begin{center}
        \includegraphics[width=\linewidth]{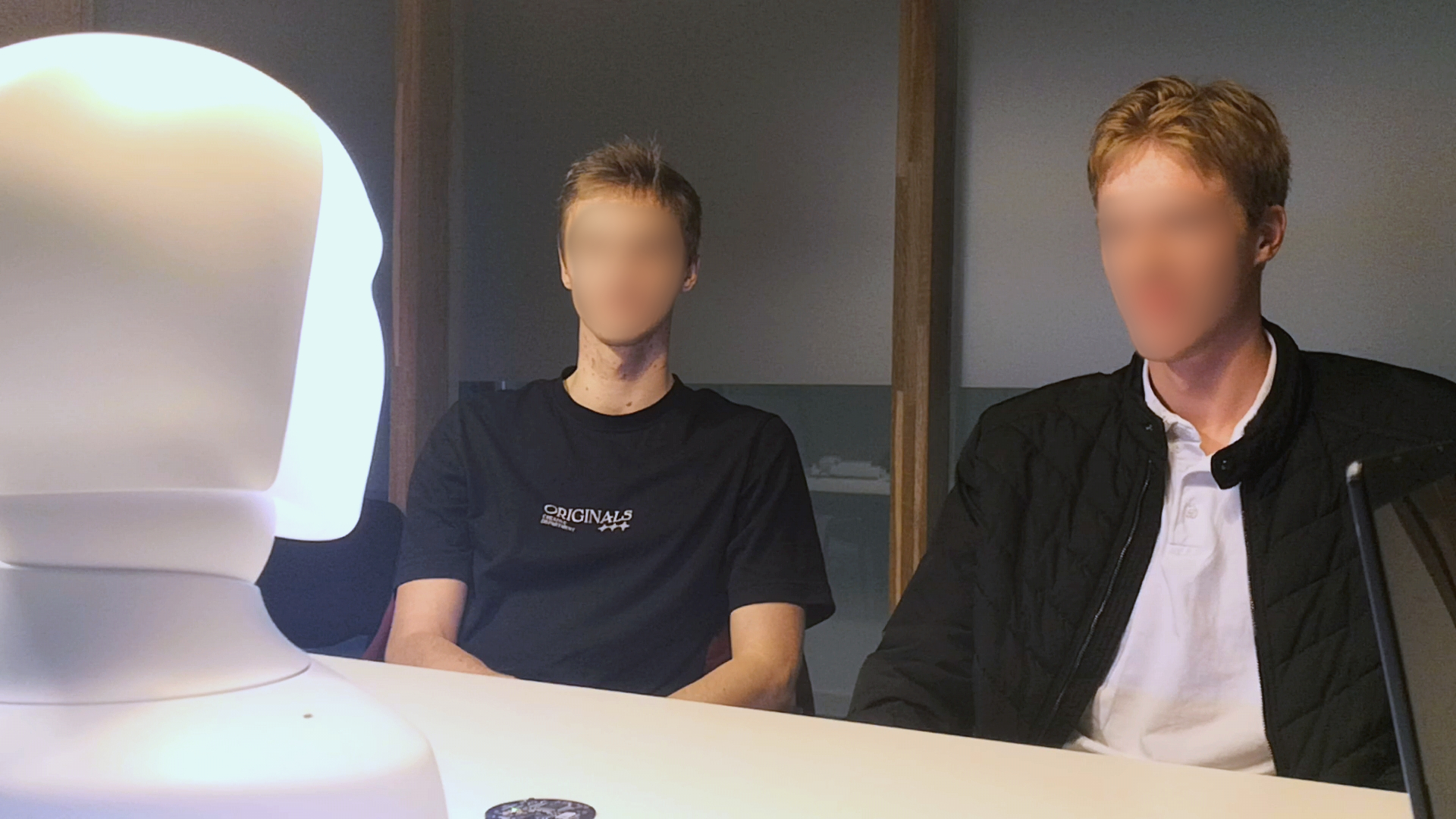}
    \end{center}
    \caption{A frame from the camera capturing the \emph{ground truth} showing two subjects interacting with the Furhat robot (on the left); the directional microphone is partially visible at the bottom.}
    \label{fig:subjects}
\end{figure}

The experiment consisted of two open-ended scenarios. In both, participants sat facing the Furhat robot, as illustrated in Figure~\ref{fig:subjects}) and all parties in the conversation spoke English. A directional microphone array and a wide-angle camera were positioned in front of the participants to support the system's perception modules and to record ground-truth audio-video data for later annotation. The order of the two scenarios was counterbalanced across participant pairs to minimise order effects.

In the \textbf{parallel scenario}, the robot alternated its attention between the two users, responding to each participant in turn. Although the conversation remained open-ended, this setting encouraged the robot to maintain two loosely independent dialogue threads. The scenario assessed the system's ability to recognise speakers, track individual conversational goals, maintain coherence with each participant's conversational direction, and manage alternating turns under overlapping demands. Note that, while turn-taking accuracy was not directly scored, we examined the success of alternating interactions and robot responsiveness through the timing and content of turn shifts. Goal-tracking was evaluated by examining whether Furhat's responses aligned with the intended conversational topic and the task of each user (e.g., discussing holiday preferences or movie genres), as coded from transcripts and dialogue intents.

In the \textbf{group scenario}, both participants and the robot engaged in a shared discussion without prescribed turn-taking rules. Furhat actively participated, contributed content, and managed its turns within a fluid, social dialogue. This condition tested the robot's ability to operate within a single conversational space, dynamically recognise addressees, and contribute coherently to an evolving group dialogue.

After each scenario, participants completed three standardised questionnaires: the Robotic Social Attributes Scale (RoSAS)~\citep{carpinella2017robotic}, which measures perceived friendliness, competence, and animacy; the Interaction Quality scale (IQ)~\citep{schmitt2015interaction}, which assesses fluency, satisfaction, and responsiveness; and the Multidimensional Measure of Trust (MDMT)~\citep{malle2021multidimensional}, which evaluates participants' trust in the robot. Each has been applied across various contexts in HRI, including robot acceptance, social engagement, and user satisfaction~\citep{roesler2022trustreview}. Questionnaires were administered separately for each scenario to obtain scenario-specific perceptions. Responses included 5-point Likert scales and optional open-ended comments, inviting participants to elaborate on positive impressions, difficulties, and desired improvements. Full questionnaire items are listed in the Supplementary Material.

\subsection{Evaluation Metrics}

Alongside the subjective questionnaires, we collected system-level performance metrics to objectively assess the system's effectiveness of the multi-party interaction system. All conversation logs, system outputs, and audio-video recordings were synchronised and manually annotated to support accurate scoring.

\textbf{Conversation length}. Each scenario lasted on average $12.4$ minutes ($SD=3.1$). The number of conversational turns and the distribution of speech across participants were extracted from transcripts to quantify engagement and participants' balance.

\textbf{Speaker recognition accuracy}. Speaker identity was estimated using two modalities: (i) Azure's speaker identification API for audio-based recognition, and (ii) Python's face recognition library for visual recognition. For both modalities, the predictions were aligned with manually annotated ground-truth labels to determine correctness.

\textbf{Addressee detection accuracy}. For every robot utterance, annotators labelled whether Furhat directed its response to the correct user based on the preceding conversational context and ground-truth recordings. Addressee labels were classified as: \textit{correct} response directed to the intended speaker, \textit{incorrect} response directed to the wrong speaker, or \textit{inclusive}  the robot intentionally addressed both users within a shared response, despite potential limitations in non-verbal alignment.

\textbf{Latency}. It was defined as the time elapsed from the end of the user's utterance transcription to the onset of the robot's verbal response. This includes multimodal processing, intent inference, and LLM generation time. Transcription delay was excluded, as it depends on the external Azure speech recognition service. The LLM was run in streaming mode to reduce response delay.

\textbf{Goal-tracking accuracy}. Dialogue transcripts were annotated to determine whether the robot's responses were coherent with the user's conversational goal at each turn. Coders assessed topic consistency, response appropriateness and the extent to which the robot maintained or shifted the ongoing conversational direction.

This combination of quantitative system metrics and subjective user evaluations provides a comprehensive perspective on the robot's multi-party conversational performance across both scenarios.

\subsection{Participants}

Thirty young adults participated in the study ($8$ female, $22$ male; $M_{\text{age}} = 22.7$, $SD = 1.88$). Participants were recruited in familiar pairs to encourage natural, comfortable interaction and to reduce social inhibition during multi-party dialogue. Participants were recruited through snowball sampling via personal and friends-of-friends networks. No compensation was provided.

Most participants ($73.3\%$) reported no prior experience with social robots, while $20.0\%$ had minimal exposure, and only $6.7\%$ had previously interacted with a social robot. All participants were fluent English speakers and reported no hearing or speech impairments. Participation was voluntary, and all procedures were approved by the Ethics Committee of Ghent University. Data were collected anonymously, and participants were free to withdraw at any time.


\section{Results and Discussion}

To evaluate the performance and perceived quality of our conversational system, we analysed quantitative metrics, participant survey responses, and qualitative feedback collected across both interaction settings. The following subsections synthesise these findings to address our three research questions: how well the system manages multi-party dialogue (RQ1), what technical barriers hinder performance (RQ2), and which aspects users identify for improvement (RQ3). Each subsection highlights distinct but interrelated dimensions of interaction quality, enabling a comprehensive discussion of system strengths, limitations, and opportunities for future refinement.

\subsection{Conversation Management Performance}

To answer RQ1 on how effectively the system managed multi-party interaction, we examined turn-taking behaviour, addressee detection, conversational coherence, and participants' perceived engagement across both interaction settings. 

\begin{figure}[!ht]
    \begin{center}
        \includegraphics[width=0.8\linewidth]{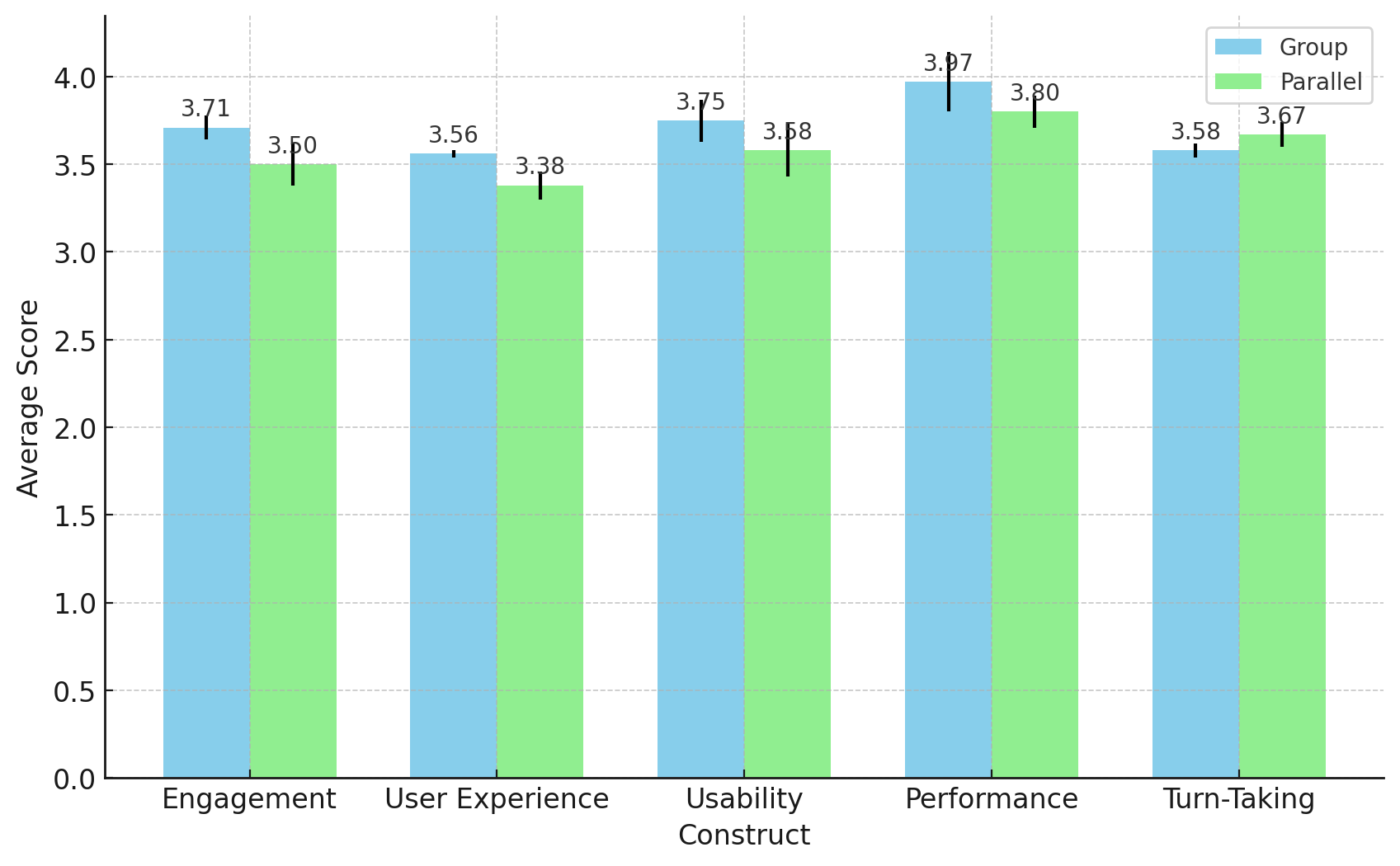}
    \end{center}
    \caption{Average ratings for each construct. Each construct has two bars -- one for the \emph{group} session and another for the \emph{parallel} -- allowing direct comparison between the two. The error bars represent the standard deviations of the ratings, providing an indication of the variability within each setting.}
\label{fig:average_rating_performance}
\end{figure}

\begin{figure}[!ht]
    \begin{center}
        \includegraphics[width=\linewidth]{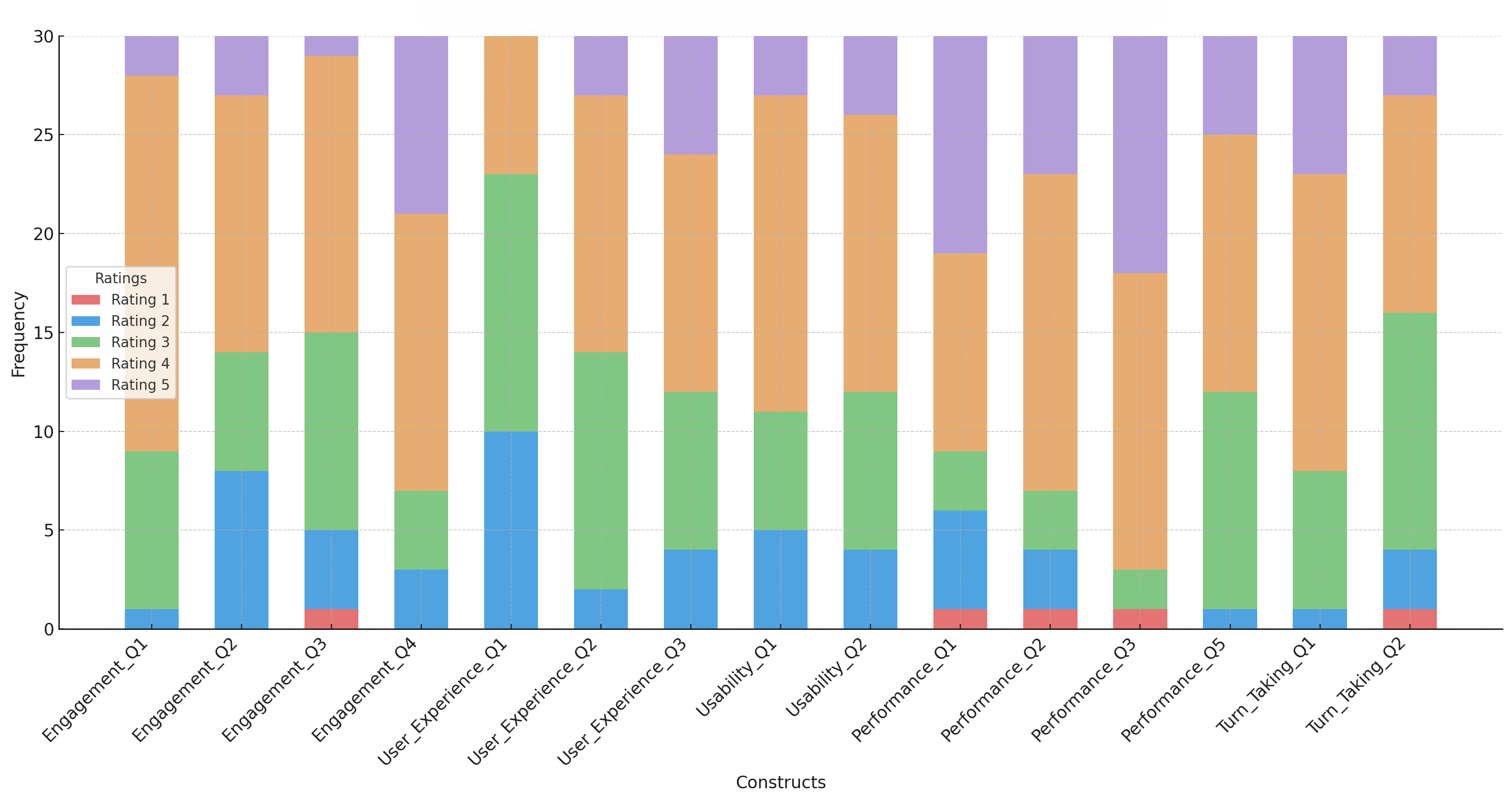}
    \end{center}
    \caption{Frequency distribution of participant ratings for each construct in the \emph{group} condition. Items are grouped by construct (e.g., Engagement, User Experience, Usability, Performance, Turn-Taking). Full questionnaire items corresponding to each code (e.g., Engagement\_Q1, Performance\_Q2) are provided in the Supplementary Material.}
    \label{fig:session_group_construct}
\end{figure}

\begin{figure}[!ht]
    \begin{center}
        \includegraphics[width=\linewidth]{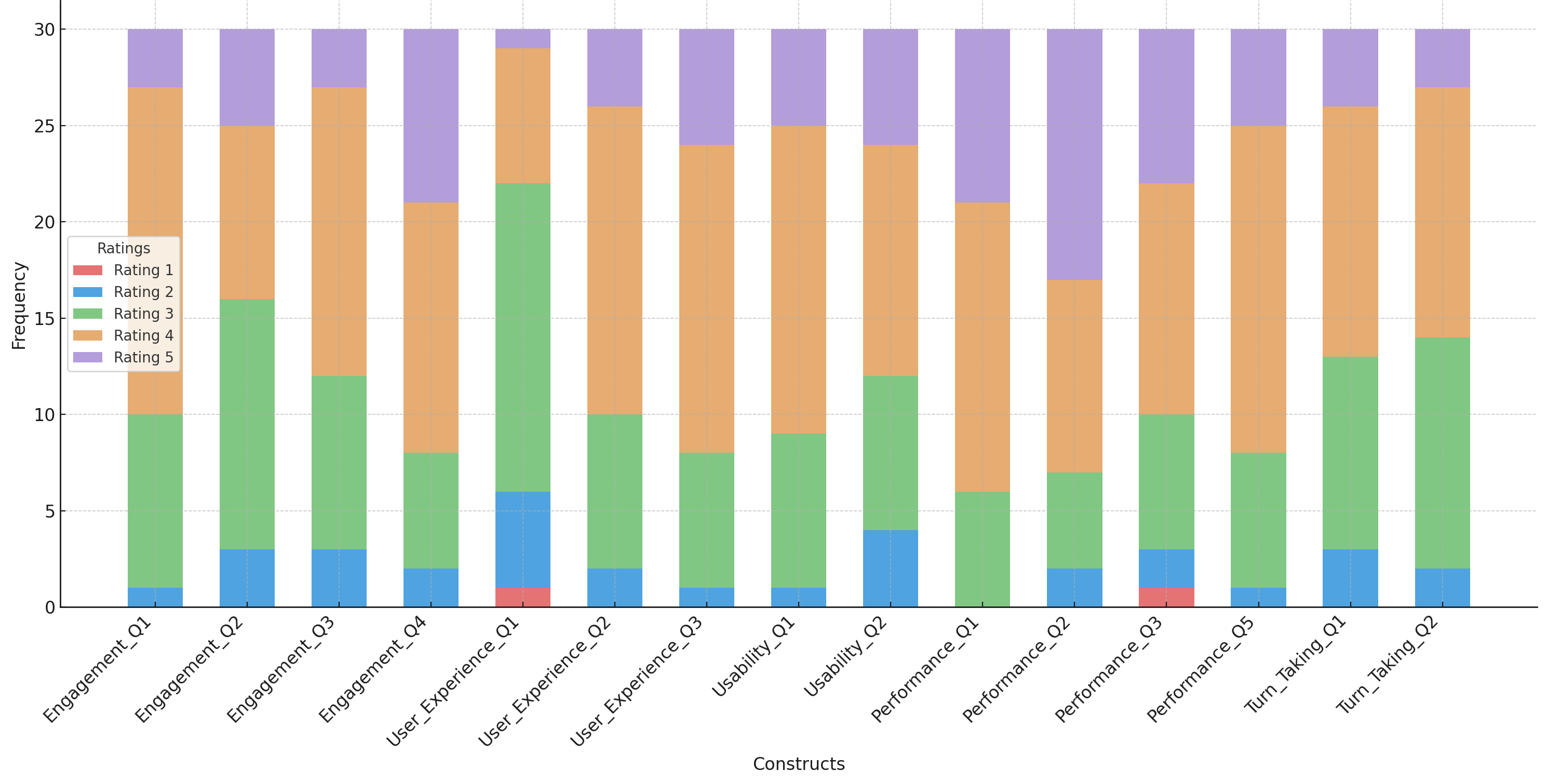}
    \end{center}
    \caption{Frequency distribution of participant ratings for each construct in the \emph{parallel} condition. Items are grouped by construct (e.g., Engagement, User Experience, Usability, Performance, Turn-Taking). Full questionnaire items corresponding to each code (e.g., Engagement\_Q1, Performance\_Q2) are provided in the Supplementary Material.}
    \label{fig:session_parallel_construct}
\end{figure}

Participants reported positive engagement in both scenarios, with group interactions slightly outperforming parallel ones (M = $3.71$, SD = $0.07$ vs M = $3.60$, SD = $0.13$), as shown in Figure~\ref{fig:average_rating_performance}. Qualitative responses support these findings; participants frequently described the group scenario as more lively, socially dynamic, and conducive to shared engagement, whereas the parallel setting was perceived as more structured and predictable. As illustrated in Figures~\ref{fig:session_group_construct} and~\ref{fig:session_parallel_construct}, group interactions also showed a higher concentration of positive ratings across several engagement- and experience-related constructs, suggesting that shared conversational spaces fostered stronger perceptions of involvement and social presence. 

Turn-taking dynamics differed noticeably between conditions. In the parallel scenario, conversational alternation between users remained relatively stable, and participants found it easier to anticipate when the robot would respond. In the group condition, however, the conversational flow became more fluid and spontaneous, leading to greater uncertainty about turn location. As shown in Table~\ref{tab:engagement_performance}, interruptions or overlapping speech occurred in $60.0\%$ of group turns compared with $56.7\%$ in the parallel settings. Participants often expressed uncertainty about their conversational turns or the intended recipient of the robot's responses, negatively influencing their interaction experience.

\begin{table}[!ht]
    \begin{center}
        \renewcommand{\arraystretch}{1.2}
        \begin{tabular}{|l|p{6cm}|l|c|c|}
            \hline
            \textbf{Construct} & \textbf{Question} & \textbf{Settings} & \textbf{Yes (\%)} & \textbf{No (\%)} \\
            \hline
            \multirow{2}{*}{Engagement} & Did you feel a sense of social presence from Furhat during the conversation? & Group & 66.7 & 33.3 \\
            & & Parallel & 50.0 & 50.0 \\
            \hline
            \multirow{2}{*}{Performance} & Did Furhat interrupt or speak over you or your partner? & Group & 60.0 & 40.0 \\
            & & Parallel & 56.7 & 43.3 \\
            \hline
            \multirow{2}{*}{Performance} & Did Furhat consistently identify the correct speaker (you or your partner) when responding to questions or comments? & Group & 73.3 & 26.7 \\
            & & Parallel & 66.7 & 33.3 \\
            \hline
        \end{tabular}
    \end{center}
    \caption{Example of survey responses regarding engagement and performance.}
    \label{tab:engagement_performance}
\end{table}

Addressee detection accuracy aligned with these observations. The system correctly directed its responses in $92.6\%$ of cases in the parallel condition but only in $79.3\%$ of cases in the group condition (see Table~\ref{tab:tab_recognition_Injection}). Participants reported that when the robot's gaze direction, head movements, and verbal cues aligned, turn-taking felt intuitive and easy to follow. One participant noted that \textit{``the robot's eye movements and head-turning made it clear when it was my turn to speak.''} However, when verbal output conflicted with nonverbal behaviour, participants experienced brief ambiguity, particularly during rapid turn-switching or overlapping utterances.

\begin{table}[!ht]
    \begin{center}
        \begin{tabular}{|l|c|c|c|}
            \hline
            \textbf{Type} & \textbf{Correct} & \textbf{Incorrect} & \textbf{Blank}\\
            \hline
            Addr. detection (group) & 79.3 (3.8 interest.) & 13.6 & 7.2 \\
            Addr. detection (parallel) & 92.6 (1.7 interest.) & 7.0 & 0.4 \\
            \hline
  	    \end{tabular}
    \end{center}
	\caption{Addressee detection accuracy (percentages).}
	\label{tab:tab_recognition_Injection}
\end{table}

Despite these inaccuracies, participants reported a stronger sense of social presence during the group setting ($66.7\%$) than during the parallel setting ($50.0\%$), as shown in Table~\ref{tab:engagement_performance}. This suggests that even when turn-taking or addressee cues were imperfect, the robot's attempts to engage both users contributed to a more inclusive and socially engaging atmosphere.

Three dialogue excerpts illustrate the strengths and limitations of the system's turn management. The first highlights misaddressing in the presence of an interruption:

\begin{quote}
    \textbf{User~A:} ``I've gotten into cycling recently, especially long weekend rides''\\
    \textbf{Robot (addressing User~A):} ``That sounds amazing! Do you ride solo or with a group?'' \\ 
    \textbf{User~B (overlapping):} ``Have you tried mountain biking?'' \\
    \textbf{Robot (misaddressing User~A):} ``Mountain biking adds a whole new challenge. What trails do you usually ride, User~A\footnote{The system used the actual name as registered during the enrolment phase.}?''
\end{quote}

Here, the system maintained the original conversational thread with User~A, which preserved coherence but failed to acknowledge User~B's overlapping contribution. This behaviour highlights the trade-off the system must make between stability and responsiveness in multi-party scenarios.

The second example demonstrates moments of inclusive addressing, where the LLM extends the scope of the response to engage both participants:

\begin{quote}
    \textbf{Robot (addressing User~A):} ``That's awesome, User~A! What do you enjoy most about roller coasters?'' \\
    \textbf{User~A:} ``I really like the ones that go really fast.'' \\
    \textbf{Robot (addressing User~A):} ``Speedy roller coasters do give a thrilling rush! User~B, are you also up for the fast-paced rides?''
\end{quote}

Participants often interpreted such turns positively when accompanied by corresponding gaze cues, viewing them as socially inclusive. However, when nonverbal cues did not shift accordingly, the inclusivity occasionally produced ambiguity about the intended addressee.

The third example shows how the system reused contextual information across participants:

 \begin{quote}
\textbf{User~A:} ``Do you want to talk about South America, for example, Peru? Because I'm going there in like 2 weeks, maybe."

\textbf{Robot (addressing User~A):} ``Peru sounds amazing! Any particular places in Peru you're excited to visit?"

\medskip

\textit{Conversation continues...}

\medskip

\textbf{User~B:} ``Hey, I'd love to discuss some travel destinations. What's the first destination that comes to your mind?"

\textbf{Robot (addressing User~B):} ``Peru comes to mind as a fascinating travel destination with its rich history and diverse landscapes. Have you ever considered visiting South America?"
\end{quote}

Here, the system later responded to User~B by drawing on relevant contextual information. This memory context helped maintain topic continuity and was perceived as coherent when it aligned with the ongoing conversational flow. 

Participants in parallel sessions particularly appreciated the clarity afforded by higher addressee accuracy and simpler conversational dynamics. Personalised addressing -- such as using names or maintaining prolonged eye contact -- was frequently mentioned as enhancing engagement. In contrast, the additional ambiguity present in group interactions sometimes weakened these positive effects, underscoring the need for improved multimodal integration and more reliable addressee identification.

Underlying these behaviours is the system's LLM-driven addressee-selection logic, which typically prioritised the most recent speaker to maintain conversational flow. When this logic aligned with gaze and head movements, the coders noted smooth transitions and coherent turn allocation. However, in cases where verbal reasoning diverged from nonverbal cues or where overlapping speech created inherent ambiguity, the system occasionally broadened its responses or misassigned turns. While inclusive addressing could be socially engaging, the absence of supporting nonverbal shifts sometimes introduced uncertainty about the intended recipient.

Taken together, these findings indicate that the system can sustain coherent multi-party interaction and demonstrates emerging strengths in inclusive engagement and topic continuity, particularly in structured parallel settings. However, precise turn-taking and reliable addressee identification remain challenging in more dynamic, overlapping conversational contexts. Improving multimodal signal integration, refining addressee detection logic, and developing more responsive turn-taking mechanisms will be essential for advancing multi-party conversational capabilities in social robots.

\subsection{Technical Barriers}

Multi-party social interaction places substantial demands on real-time perception and response generation. Across both conditions, several technical behaviours consistently shaped the naturalness and fluidity of the interaction. Addressing RQ2, three factors emerged as the most influential: (i) the difficulty of audio-based speaker identification during overlapping or rapid turn-taking, (ii) inconsistencies in multimodal cue integration, and (iii) response latency.

Audio-based speaker recognition emerged as a notably problematic component across both interaction settings, although its limitations were more pronounced during group interaction. As shown in Table~\ref{tab_recognition_performance}, only $18.4\%$ of group utterances and $26.8\%$ of parallel utterances were correctly attributed, with most remaining instances classified as unrecognised. These low percentages do not indicate malfunction but reflect the inherent difficulty of recognising speakers in multi-party environments characterised by short interjections, overlapping speech, and rapid turn transitions. Participants frequently attributed breakdowns in interaction to these audio recognition limitations, with comments such as: \textit{``I had to repeat myself several times because it didn't know who was talking.''} These observations reinforce that improving robustness under conditions of crosstalk and simultaneous contributions is essential for natural multi-party dialogue.

\begin{table}[!ht]
    \begin{center}
        \begin{tabular}{|l|c|c|c|}
            \hline
            \textbf{Type} & \textbf{Correct} & \textbf{Incorrect} & \textbf{Blank}\\
            \hline
            Voice recognition (group) & 18.4 & 4.5 & 77.0\\
            Voice recognition (parallel) & 26.8 & 0.4 & 72.8\\
            \hline
            Face recognition (group) & 80.0 & 19.5 & 0.5 \\
            Face recognition (parallel) & 94.7 & 4.2 & 1.1\\
            \hline
  	    \end{tabular}
    \end{center}
	\caption{Voice and face recognition accuracy (percentages). Note that \textit{Correct recognition} was defined as a match between the system's prediction and the speaker's identity as verified through annotated video ground truth; \textit{incorrect recognition} reflected mismatches, while \textit{blank} denotes instances with no speaker ID detected or recognised. Here, \emph{voice recognition} refers to real-time speaker identification (matching utterances to known speaker profiles), not diarisation or speech-to-text transcription.}
	\label{tab_recognition_performance}
\end{table}

In contrast to audio, visual speaker detection performed reliably across conditions, achieving $80.0\%$ accuracy in group interactions and $94.7\%$ accuracy in parallel interactions. Annotators pointed out that when voice recognition failed, visual identification often enabled the robot to maintain coherence in its response. However, when audio and visual predictions diverged, the system occasionally produced misaligned behaviours, such as looking at one participant while replying to the other. As one participant noted: \textit{``At times, Furhat clearly looked at me but responded to my partner, causing confusion about whom it was addressing.''} These mismatches highlight the need for confidence-weighted multimodal fusion that can dynamically prioritise the more reliable modality -- typically vision -- during periods of overlapping speech or rapid turn-switching.

Latency constituted the second major barrier to fluid interaction. The average response delay was $1.35$ seconds (SD = $0.55$), occasionally reaching nearly five seconds. While such delays are acceptable in dyadic conversation, they disrupted group dialogues where participants often shifted topics or resumed speaking before the robot responded. Participants reported that the robot occasionally felt slow or detached: \textit{``The delay made the conversation feel sluggish,''} and \textit{``It broke the rhythm.''} Analysis of system logs indicates that LLM generation time was the largest contributor to this delay (mean $0.76$ seconds). This suggests that future iterations would benefit from lightweight or predictive generation strategies, particularly for fast-paced social interactions.

These technical behaviours were directly reflected in post-interaction questionnaire responses (Table~\ref{tab:engagement_performance}). Reports of being interrupted or misidentified were more frequent in the group condition, where rapid conversational shifts amplified the combined effects of delays and recognition errors. Participants in these sessions often experienced reduced naturalness, increased conversational overlap, and disrupted turn-taking cues. In contrast, the parallel condition, characterised by clearer conversational structure and fewer simultaneous contributions, mitigated some of the negative impact of latency. Participants in parallel sessions were better able to accommodate delayed responses and experienced fewer compounded failures between recognition and timing.

Across both settings, audio recognition fragility, modality conflicts, and LLM-induced delays emerged as the main limiting factors of multi-party performance. These findings highlight the importance of developing more adaptive multimodal fusion strategies, strengthening speaker tracking under overlapping speech, and reducing response latency to support natural, real-time social interaction. Addressing these challenges will be crucial for enabling LLM-driven robots to operate effectively in multi-party settings.

\subsection{User Perceptions and Design Insights}

Participants' reflections offered a complementary perspective to the objective performance metrics, revealing how the system's behaviours were interpreted, tolerated, or valued during interaction. While overall engagement ratings were positive across both conditions, open-ended comments indicated specific areas where the interaction could be made more natural, personal, and intuitive.

Across both interaction settings, participants consistently expressed a preference for greater personalisation and contextual awareness in the robot's behaviour. Although mean ratings for social perception remained above $3.5$ in both scenarios, open-ended feedback revealed that several responses felt generic or insufficiently tailored to the individual. Around $16.7\%$ of participants explicitly stated that the robot did not make use of information they had previously shared. One participant remarked, \textit{``It would have been nice if Furhat remembered something I said earlier, bringing it back later to personalise the interaction.''} 

The system already included a memory mechanism capable of storing user information, yet the language model did not consistently surface this knowledge in its responses. The participants, therefore, perceived the robot as lacking memory despite the underlying capability being present. This gap illustrates a wider design challenge in LLM-driven systems. Memory must not only exist at the architectural level but must also be expressed in ways that are perceptible and meaningful to users, such as through timely references to earlier statements or personalised conversational follow-ups.

Participants frequently commented on the formal tone of the robot's utterances. Although $76.7\%$ rated its conversational appropriateness as natural, many noted that its replies, while accurate, often felt too polished or scripted (Table~\ref{tab:yes_no_summary}). Descriptions such as too perfect or overly formal appeared repeatedly in feedback. Participants explained that these characteristics reduced the spontaneity of the interaction and added to the impression of conversational rigidity. This experience was often intensified by response delays, which made the robot feel slow or detached and diminished the perceived naturalness of turn-taking.

\begin{table}[!ht]
    \begin{center}
        \begin{tabular}{|p{10cm}|c|c|}
            \hline
            \textbf{Question} & \textbf{Yes (\%)} & \textbf{No (\%)} \\
            \hline
            Did Furhat successfully follow the flow of the conversation between you and your partner? & 96.67 & 3.33 \\
            Did Furhat contribute to the conversation in a natural way? & 76.67 & 23.33 \\
            Did Furhat effectively provide prompts that encouraged further discussion? & 70.00 & 30.00 \\
            In individual conversations, did Furhat differentiate effectively between the speaker and the partner? & 86.67 & 13.33 \\
            Did Furhat appropriately respond to personal preferences in the separate conversation? & 83.33 & 16.67 \\
            Did Furhat mistakenly involve the other person in your conversation or direct a response incorrectly? & 33.33 & 66.67 \\
            Was Furhat's response length appropriate to allow for a balanced flow of the conversation? & 90.00 & 10.00 \\
            \hline
        \end{tabular}
    \end{center}
    \caption{Summary of responses from participants on Furhat's conversational skills.}
    \label{tab:yes_no_summary}
\end{table}

Turn-taking cues were another recurring theme in participant reflections. While quantitative results indicated moderate success in identifying speakers, many participants reported difficulty in knowing when to speak or when a response was intended for them. This uncertainty was especially common in the group setting, where conversational dynamics were more fluid. Participants recommended more explicit nonverbal cues, such as clearer gaze shifts, more distinct head turns, or small gestures to signal turn boundaries. Comments emphasised that even subtle adjustments in these behaviours would greatly improve conversational clarity.

User reflections on recognition issues closely paralleled the quantitative findings.
Participants frequently noted moments when they felt unheard or misidentified, often linking these incidents to reduced engagement or conversational flow. Statements like \textit{``Sometimes Furhat didn't respond because it couldn't hear me clearly,''} or \textit{``I had to repeat myself several times,''} reflect how recognition errors manifest subjectively as conversational breakdowns. Interestingly, users did not typically distinguish between audio and visual recognition; instead, mismatches such as the robot looking at one person while addressing another were experienced simply as inconsistency, reinforcing the importance of more transparent and conflict-resolving multimodal fusion.

Alongside these recommendations, several participants emphasised aspects of the system they found particularly effective. The robot's nonverbal behaviour, especially its eye contact, head orientation, and subtle facial expressions, was frequently described as engaging and lifelike. Participants also noted that the robot maintained conversational coherence even when recognition errors occurred, which they attributed to its consistent topic-following and contextual reasoning. These observations suggest that users perceive the system as socially capable and expressive, even when technical limitations affect timing or addressee accuracy.

Taken together, these findings provide a response to RQ3. Users identified several areas where system behaviour could be improved, most notably in personalisation, spontaneity of language, clarity of turn-taking cues, consistency across modalities, and response timing. At the same time, they recognised strengths in nonverbal expressiveness and overall social presence. These insights outline concrete design priorities for future iterations of multi-party conversational robots. 

\subsection{Limitations and Future Work}

While the study provides clear insights into the capabilities and challenges of LLM-driven multi-party interaction, several limitations should be acknowledged. First, the evaluation involved a modest sample size of $30$ young adults, which constrains the generalisability of the findings and does not capture the broader range of conversational styles found in real-world settings. 

Second, the interactions were conducted under controlled laboratory conditions. Although the group scenario introduced natural overlap and spontaneity, the structured environment may not fully reflect the complexity and unpredictability of everyday multi-party encounters.

Third, several performance outcomes are closely tied to the specific architectural choices made (e.g, relying on Azure's speaker identification, GPT-3.5 for response generation). Different recognition pipelines or more recent LLMs may yield substantially different results. Similarly, the relatively short duration of the interactions limits conclusions about long-term adaptation, multi-session memory use, or behaviour over extended deployments.  Future research should therefore explore larger group settings, more diverse user populations, and longer-term engagements to assess how LLM-driven conversational architectures scale and adapt beyond controlled experimental setups.

\section{Conclusions}

This work presented and evaluated a multimodal conversational architecture enabling a social robot to engage in open-ended dialogue with two users simultaneously. Through a controlled two-condition study, we examined how the system coordinated turn-taking, identified speakers, selected addressees, and responded to rapidly shifting conversational dynamics.

Our findings show that the system can sustain coherent multi-party interaction, particularly in structured parallel settings where addressee accuracy and topic continuity remained high. In more fluid group interactions, the robot demonstrated emerging strengths, such as inclusive addressing and robust visual recognition, while also revealing the challenges posed by overlapping speech, modality conflicts, and response latency. User reflections highlighted both the value of the robot's social expressiveness and the need for clearer turn-taking cues, stronger personalisation, and more reliable multimodal grounding.

Overall, the study provides a detailed view of the capabilities and limitations of current LLM-driven conversational systems in multi-party contexts. The findings point to several technical priorities for future work, including more adaptive multimodal fusion, improved handling of overlapping speech, and faster response generation to support smoother conversational flow. Extending the system to larger groups, longer-term interactions, and newer language and perception models will help advance the development of socially capable robots that can operate effectively in natural, dynamic, real-world settings.

\section*{Funding}

Funded by the Horizon Europe VALAWAI project (grant agreement number 101070930), Bijzonder Onderzoeksfonds (BOF) of Ghent University (grant BOF22/DOC/235) and the Flanders AI Research 2 project.

\bibliographystyle{ACM-Reference-Format}
\bibliography{main}

@inproceedings{richter2016you,
  title={Are you talking to me? Improving the robustness of dialogue systems in a multi party HRI scenario by incorporating gaze direction and lip movement of attendees},
  author={Richter, Viktor and Carlmeyer, Birte and Lier, Florian and Meyer zu Borgsen, Sebastian and Schlangen, David and Kummert, Franz and Wachsmuth, Sven and Wrede, Britta},
  booktitle={Proceedings of the fourth international conference on human agent interaction},
  pages={43--50},
  year={2016}
}

@inproceedings{paetzel2023improving,
  title={Improving a Robot's Turn-Taking Behavior in Dynamic Multiparty Interactions},
  author={Paetzel-Pr{\"u}smann, Maike and Kennedy, James},
  booktitle={Companion of the 2023 ACM/IEEE International Conference on Human-Robot Interaction},
  pages={411--415},
  year={2023}
}

@inproceedings{foster2012two,
  title={Two people walk into a bar: Dynamic multi-party social interaction with a robot agent},
  author={Foster, Mary Ellen and Gaschler, Andre and Giuliani, Manuel and Isard, Amy and Pateraki, Maria and Petrick, Ronald PA},
  booktitle={Proceedings of the 14th ACM international conference on Multimodal interaction},
  pages={3--10},
  year={2012}
}

@inproceedings{carpinella2017robotic,
  title={The robotic social attributes scale (rosas) development and validation},
  author={Carpinella, Colleen M and Wyman, Alisa B and Perez, Michael A and Stroessner, Steven J},
  booktitle={Proceedings of the 2017 ACM/IEEE International Conference on human-robot interaction},
  pages={254--262},
  year={2017}
}

@article{cifuentes2020social,
  title={Social robots in therapy and care},
  author={Cifuentes, Carlos A and Pinto, Maria J and C{\'e}spedes, Nathalia and M{\'u}nera, Marcela},
  journal={Current Robotics Reports},
  volume={1},
  pages={59--74},
  year={2020},
  publisher={Springer}
}

@article{roesler2022trustreview,
  title={Measuring trust in social robots: A systematic review},
  author={Roesler, Eileen and Schmid, Julia and Eyssel, Friederike},
  journal={ACM Transactions on Human-Robot Interaction (THRI)},
  volume={11},
  number={1},
  pages={3:1--3:35},
  year={2022},
  publisher={ACM}
}

@article{obaigbena2024ai,
  title={AI and human-robot interaction: A review of recent advances and challenges},
  author={Obaigbena, Alexander and Lottu, Oluwaseun Augustine and Ugwuanyi, Ejike David and Jacks, Boma Sonimitiem and Sodiya, Enoch Oluwademilade and Daraojimba, Obinna Donald},
  journal={GSC Advanced Research and Reviews},
  volume={18},
  number={2},
  pages={321--330},
  year={2024},
  publisher={GSC Advanced Research and Reviews}
}

@article{schobel2024charting,
  title={Charting the evolution and future of conversational agents: A research agenda along five waves and new frontiers},
  author={Sch{\"o}bel, Sofia and Schmitt, Anuschka and Benner, Dennis and Saqr, Mohammed and Janson, Andreas and Leimeister, Jan Marco},
  journal={Information Systems Frontiers},
  volume={26},
  number={2},
  pages={729--754},
  year={2024},
  publisher={Springer}
}

@article{wolfert2022review,
  title={A review of evaluation practices of gesture generation in embodied conversational agents},
  author={Wolfert, Pieter and Robinson, Nicole and Belpaeme, Tony},
  journal={IEEE Transactions on Human-Machine Systems},
  volume={52},
  number={3},
  pages={379--389},
  year={2022},
  publisher={IEEE}
}

@inproceedings{aneja2021understanding,
  title={Understanding conversational and expressive style in a multimodal embodied conversational agent},
  author={Aneja, Deepali and Hoegen, Rens and McDuff, Daniel and Czerwinski, Mary},
  booktitle={Proceedings of the 2021 CHI conference on human factors in computing systems},
  pages={1--10},
  year={2021}
}

@article{gibson2000seizing,
  title={Seizing the moment: The problem of conversational agency},
  author={Gibson, David R},
  journal={Sociological Theory},
  volume={18},
  number={3},
  pages={368--382},
  year={2000},
  publisher={SAGE Publications Sage CA: Los Angeles, CA}
}

@inproceedings{verhelst2024adaptive,
  title={Adaptive Second Language Tutoring Using Generative AI and a Social Robot},
  author={Verhelst, Eva and Janssens, Ruben and Demeester, Thomas and Belpaeme, Tony},
  booktitle={Companion of the 2024 ACM/IEEE International Conference on Human-Robot Interaction},
  pages={1080--1084},
  year={2024}
}

@article{belpaeme2018social,
  title={Social robots for education: A review},
  author={Belpaeme, Tony and Kennedy, James and Ramachandran, Aditi and Scassellati, Brian and Tanaka, Fumihide},
  journal={Science robotics},
  volume={3},
  number={21},
  pages={eaat5954},
  year={2018},
  publisher={American Association for the Advancement of Science}
}

@article{kyrarini2021survey,
  title={A survey of robots in healthcare},
  author={Kyrarini, Maria and Lygerakis, Fotios and Rajavenkatanarayanan, Akilesh and Sevastopoulos, Christos and Nambiappan, Harish Ram and Chaitanya, Kodur Krishna and Babu, Ashwin Ramesh and Mathew, Joanne and Makedon, Fillia},
  journal={Technologies},
  volume={9},
  number={1},
  pages={8},
  year={2021},
  publisher={MDPI}
}

@article{broadbent2023enhancing,
  title={Enhancing social connectedness with companion robots using AI},
  author={Broadbent, Elizabeth and Billinghurst, Mark and Boardman, Samantha G and Doraiswamy, P Murali},
  journal={Science Robotics},
  volume={8},
  number={80},
  pages={eadi6347},
  year={2023},
  publisher={American Association for the Advancement of Science}
}

@inproceedings{giudici2024delivering,
  title={Delivering Green Persuasion strategies with a conversational agent: a pilot study},
  author={Giudici, Mathyas and Abbo, Giulio Antonio and Crovari, Pietro and Garzotto, Franca},
  booktitle={57th Hawaii International Conference on System Sciences},
  pages={811--820},
  year={2024}
}

@inproceedings{addlesee2024multi,
  title={A multi-party conversational social robot using LLMS},
  author={Addlesee, Angus and Cherakara, Neeraj and Nelson, Nivan and Hern{\'a}ndez Garc{\'\i}a, Daniel and Gunson, Nancie and Siei{\'n}ska, Weronika and Romeo, Marta and Dondrup, Christian and Lemon, Oliver},
  booktitle={Companion of the 2024 ACM/IEEE International Conference on Human-Robot Interaction},
  pages={1273--1275},
  year={2024}
}

@article{chang2024survey,
  title={A survey on evaluation of large language models},
  author={Chang, Yupeng and Wang, Xu and Wang, Jindong and Wu, Yuan and Yang, Linyi and Zhu, Kaijie and Chen, Hao and Yi, Xiaoyuan and Wang, Cunxiang and Wang, Yidong and others},
  journal={ACM Transactions on Intelligent Systems and Technology},
  volume={15},
  number={3},
  pages={1--45},
  year={2024},
  publisher={ACM New York, NY}
}

@incollection{malle2021multidimensional,
  title={A multidimensional conception and measure of human-robot trust},
  author={Malle, Bertram F and Ullman, Daniel},
  booktitle={Trust in human-robot interaction},
  pages={3--25},
  year={2021},
  publisher={Elsevier}
}

@article{schmitt2015interaction,
  title={Interaction quality: assessing the quality of ongoing spoken dialog interaction by experts—and how it relates to user satisfaction},
  author={Schmitt, Alexander and Ultes, Stefan},
  journal={Speech Communication},
  volume={74},
  pages={12--36},
  year={2015},
  publisher={Elsevier}
}

@inproceedings{moujahid2022multi,
  title={Multi-party interaction with a robot receptionist},
  author={Moujahid, Meriam and Hastie, Helen and Lemon, Oliver},
  booktitle={2022 17th ACM/IEEE International Conference on Human-Robot Interaction (HRI)},
  pages={927--931},
  year={2022},
  organization={IEEE}
}

@article{zarkowski2019multi,
  title={Multi-party turn-taking in repeated human--robot interactions: an interdisciplinary evaluation},
  author={{\.Z}arkowski, Mateusz},
  journal={International Journal of Social Robotics},
  volume={11},
  number={5},
  pages={693--707},
  year={2019},
  publisher={Springer}
}

@inproceedings{murali2023improving,
  title={Improving multiparty interactions with a robot using large language models},
  author={Murali, Prasanth and Steenstra, Ian and Yun, Hye Sun and Shamekhi, Ameneh and Bickmore, Timothy},
  booktitle={Extended Abstracts of the 2023 CHI Conference on Human Factors in Computing Systems},
  pages={1--8},
  year={2023}
}

@inproceedings{al2012furhat,
  title={Furhat: a back-projected human-like robot head for multiparty human-machine interaction},
  author={Al Moubayed, Samer and Beskow, Jonas and Skantze, Gabriel and Granstr{\"o}m, Bj{\"o}rn},
  booktitle={Cognitive Behavioural Systems: COST 2102 International Training School, Dresden, Germany, February 21-26, 2011, Revised Selected Papers},
  pages={114--130},
  year={2012},
  organization={Springer}
}

@article{skantze2021turn,
  title={Turn-taking in conversational systems and human-robot interaction: a review},
  author={Skantze, Gabriel},
  journal={Computer Speech \& Language},
  volume={67},
  pages={101178},
  year={2021},
  publisher={Elsevier}
}

@inproceedings{chidambaram2012designing,
  title={Designing persuasive robots: how robots might persuade people using vocal and nonverbal cues},
  author={Chidambaram, Vijay and Chiang, Yueh-Hsuan and Mutlu, Bilge},
  booktitle={Proceedings of the seventh annual ACM/IEEE international conference on Human-Robot Interaction},
  pages={293--300},
  year={2012}
}

@article{mori2012uncanny,
  title={The uncanny valley [from the field]},
  author={Mori, Masahiro and MacDorman, Karl F and Kageki, Norri},
  journal={IEEE Robotics \& automation magazine},
  volume={19},
  number={2},
  pages={98--100},
  year={2012},
  publisher={IEEE}
}

@article{park2022review,
  title={A review of speaker diarization: Recent advances with deep learning},
  author={Park, Tae Jin and Kanda, Naoyuki and Dimitriadis, Dimitrios and Han, Kyu J and Watanabe, Shinji and Narayanan, Shrikanth},
  journal={Computer Speech \& Language},
  volume={72},
  pages={101317},
  year={2022},
  publisher={Elsevier}
}

@article{adjabi2020past,
  title={Past, present, and future of face recognition: A review},
  author={Adjabi, Insaf and Ouahabi, Abdeldjalil and Benzaoui, Amir and Taleb-Ahmed, Abdelmalik},
  journal={Electronics},
  volume={9},
  number={8},
  pages={1188},
  year={2020},
  publisher={MDPI}
}

@inproceedings{moriya2012estimation,
  title={Estimation of conversational activation level during video chat using turn-taking information.},
  author={Moriya, Yurie and Tanaka, Takahiro and Miyajima, Toshimitu and Fujita, Kinya},
  booktitle={Proceedings of the 10th asia pacific conference on Computer human interaction},
  pages={289--298},
  year={2012}
}

@article{nigro2024social,
  title={Social Group Human-Robot Interaction: A Scoping Review of Computational Challenges},
  author={Nigro, Massimiliano and Akinrintoyo, Emmanuel and Salomons, Nicole and Spitale, Micol},
  journal={arXiv e-prints},
  pages={arXiv--2412},
  year={2024}
}

@article{ghiglino2021mind,
  title={Mind the eyes: artificial agents’ eye movements modulate attentional engagement and anthropomorphic attribution},
  author={Ghiglino, Davide and Willemse, Cesco and De Tommaso, Davide and Wykowska, Agnieszka},
  journal={Frontiers in Robotics and AI},
  volume={8},
  pages={642796},
  year={2021},
  publisher={Frontiers Media SA}
}

@article{mehmood2024embracing,
  title={Embracing digital companions: Unveiling customer engagement with anthropomorphic AI service robots in cross-cultural context},
  author={Mehmood, Khalid and Kautish, Pradeep and Shah, Tejas R},
  journal={Journal of Retailing and Consumer Services},
  volume={79},
  pages={103825},
  year={2024},
  publisher={Elsevier}
}

@inproceedings{skantze-turn-taking,
author = {Skantze, Gabriel and Irfan, Bahar},
title = {Applying General Turn-taking Models to Conversational Human-Robot Interaction},
year = {2025},
publisher = {IEEE Press},
booktitle = {Proceedings of the 2025 ACM/IEEE International Conference on Human-Robot Interaction},
pages = {859–868},
numpages = {10},
location = {Melbourne, Australia},
series = {HRI '25}
}

@INPROCEEDINGS{10973830,
  author={Abbo, Giulio Antonio and Belpaeme, Tony},
  booktitle={2025 20th ACM/IEEE International Conference on Human-Robot Interaction (HRI)}, 
  title={I Was Blind but Now I See: Implementing Vision-Enabled Dialogue in Social Robots}, 
  year={2025},
  volume={},
  number={},
  pages={1176-1180},
  doi={10.1109/HRI61500.2025.10973830}}

@inproceedings{addlesee-etal-2023-multi,
    title = "Multi-party Goal Tracking with {LLM}s: Comparing Pre-training, Fine-tuning, and Prompt Engineering",
    author = "Addlesee, Angus  and
      Siei{\'n}ska, Weronika  and
      Gunson, Nancie  and
      Hernandez Garcia, Daniel  and
      Dondrup, Christian  and
      Lemon, Oliver",
    editor = "Stoyanchev, Svetlana  and
      Joty, Shafiq  and
      Schlangen, David  and
      Dusek, Ondrej  and
      Kennington, Casey  and
      Alikhani, Malihe",
    booktitle = "Proceedings of the 24th Annual Meeting of the Special Interest Group on Discourse and Dialogue",
    month = sep,
    year = "2023",
    address = "Prague, Czechia",
    publisher = "Association for Computational Linguistics",
    url = "https://aclanthology.org/2023.sigdial-1.22/",
    doi = "10.18653/v1/2023.sigdial-1.22",
    pages = "229--241",
}

@article{branigan2006perspectives,
  title={Perspectives on multi-party dialogue},
  author={Branigan, Holly},
  journal={Research on Language and Computation},
  volume={4},
  pages={153--177},
  year={2006},
  publisher={Springer}
}

@article{clark1992dealing,
  title={Dealing with overhearers},
  author={Clark, Herbert H and Schaefer, Edward F},
  journal={Arenas of language use},
  pages={248--297},
  year={1992},
  publisher={University of Chicago Press Chicago}
}

@article{clark1982hearers,
  title={Hearers and speech acts},
  author={Clark, Herbert H and Carlson, Thomas B},
  journal={Language},
  volume={58},
  number={2},
  pages={332--373},
  year={1982},
  publisher={Linguistic Society of America}
}
\clearpage
\appendix

\section{Implementation Details}
\label{app:implementation}

\subsection{Architecture}

\begin{figure}[!ht]
    \begin{center}
        \includegraphics[width=0.7\linewidth]{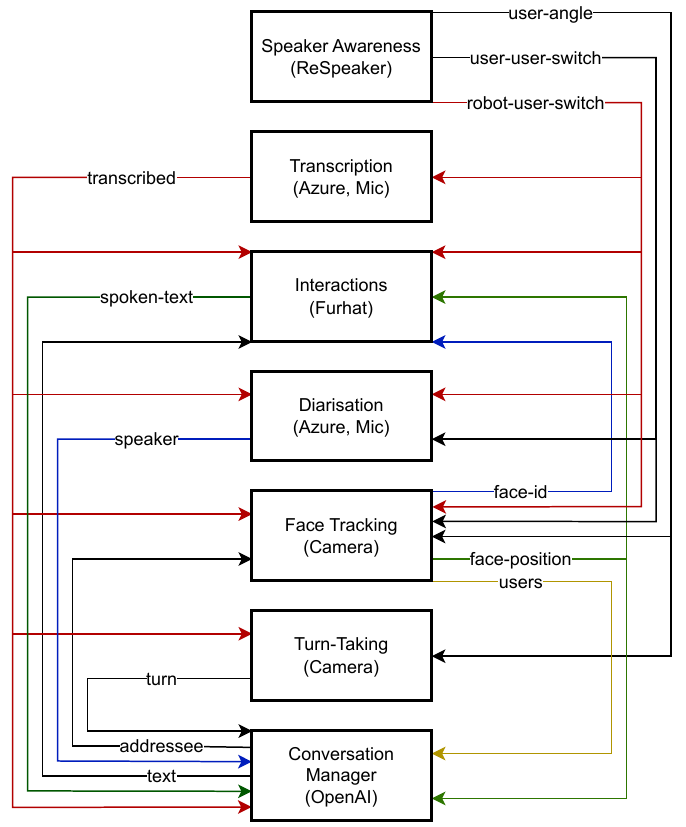}
    \end{center}
    \caption{Overview of the system's components and their connections. External components are between parenthesis: the ReSpeaker microphone for direction of arrival, or ``Mic'' for the audio stream, the Furhat robot or the Camera feed from the robot, and the external services used. Colours are only to help distinguish between the lines.}
    \label{fig:components}
\end{figure}

In the following section, we provide additional details on the architecture and implementation of the system.
For each module, we specify its function, what the component publishes and to what it subscribes. An overview is reported in Figure~\ref{fig:components}.
Finally, when applicable, we provide a reflection with the assumptions for this component, its limitations, and future work.

\subsubsection{Speaker Awareness Module} 

\paragraph{Function.}
This module detects the direction of arrival of the user's voice. Using this information, it detects whether there might have been a turn switch between two users or between a user and the robot.
For the direction of arrival, the degrees are calculated from the perspective of the robot, having 0 degrees at the right, increasing counter-clockwise.
Speech coming from between 180 and 360 degrees is interpreted as coming from the robot.
The microphone used is a ReSpeaker USB Mic Array\footnote{\url{https://wiki.seeedstudio.com/ReSpeaker-USB-Mic-Array/}} positioned in front of the robot.
This device has four microphones and can run speech algorithms on-chip.
It is controlled through a Python script following the documentation of the manufacturer.

\paragraph{Pub: user-angle.} Every 100 ms, this publishes an integer with the direction of arrival of the audio, unless it comes from the robot's direction.

\paragraph{Pub: user-user-switch.} Every 100 ms, this checks whether the direction of arrival has changed by more than 20 degrees with respect to the previous measurement.
If so, an event is fired.
The measurement ignores audio coming from the robot's direction.

\paragraph{Pub: robot-user-switch.}
Every 100 ms, fire an event if the direction of arrival has changed from the user to the robot or vice versa.
The event details whether the turn switched to or from the robot.

\paragraph{Reflection.}
This component is implemented with two assumptions: first, that no one is standing behind or on the side of the robot.
Second, the participants have at least a 20-degree angle between them and the microphone.
Further testing is needed to assess the system's behaviour when these requirements are not met, as in principle it can rely on information from other components to overcome these limitations.
The turn-switch logic should be moved to another component, as it can benefit from input from other components.
Detecting the robot's voice might take advantage of information from the Interactions Module.
The ReSpeaker hardware can run small algorithms to improve the signal and provide additional information, which could improve the system's performance.

\subsubsection{Transcription Module} 

\paragraph{Function.}
Transcribes text and identifies participants through the conversation transcriber from Azure Cognitive speech services.
The ID assigned to the voices is not consistent across interactions, it only differentiates between the voices detected.

\paragraph{Pub: transcribed.}
When a result from the transcriber is received, this posts the transcribed text, its start time, and the ID of the speaker.


\paragraph{Sub: Speaker Awareness -- robot-user-switch.}
Pause the transcription when the turn is switching to the robot; resume it when the turn is switching to a user.

\paragraph{Reflection.}
In the future, we plan to experiment with different speech-to-text engines, including Whisper models, which showed promising performance and support multiple languages.

\subsubsection{Interactions Module} 

\paragraph{Function.}
This component interfaces with the Furhat robot through the Python remote API provided by the manufacturer. 
It is used for speaking utterances and controlling the gaze direction.
It handles interruptions, as it stops the robot's speech when the users start talking.
Finally, it controls the robot's LEDs, which are turned on when the robot is listening.

\paragraph{Pub: spoken-text.}
Publishes the sentence that has been spoken by the robot.

\paragraph{Sub: Speaker Awareness -- robot-user-switch.}
If the user starts talking while the robot is talking, the robot stops.
If no speech from the user is detected for at least 1.5 seconds, the robot continues talking from where it stopped.

\paragraph{Sub: Transcription -- transcribed.}
The robot shows a feedback gesture -- \emph{brow raise} -- in response to a transcribe event.


\paragraph{Sub: Conversation Manager -- text.}
When this component receives a sentence that the robot should say, the robot's LEDs are turned off.
Then, this component sends the received text to the robot's text-to-speech module.
The text is not sent if an interruption happened less than two seconds before, or if a person was talking (i.e., a transcription was received) less than one second before.
Finally, the LEDs are turned back on to signal that the robot is listening again.


\paragraph{Sub: Face Tracking -- face-position.}
When a user is talking and their face's position is known, this instructs the robot to look at the direction specified.

\paragraph{Sub: Face Tracking -- face-id.}
Instructs the robot to look at the user with the specified ID.
The ID is assigned and handled by the Furhat robot.

\paragraph{Reflection.}
The logic detecting interruptions should be moved outside of this component.
In addition, more facial expressions could be included as feedback to the user.

\subsubsection{Diarisation Module} 

\paragraph{Function.}
This component records short pieces of the conversation and uses Azure speaker recognition's text-independent identification service to assign a unique ID to the voices being recorded.
This is slower than the transcription but provides a unique ID.
Before the interaction, users are enrolled into the system, by asking them to read out loud a short paragraph.

\paragraph{Pub: speaker.}
Publish an event when new text is associated with a user.

\paragraph{Sub: Transcription -- transcribed.}
When the text of a transcription is received, this component associates it with the user that has been identified by the Azure services. 
Then, the audio recording is stopped if it is running.

\paragraph{Sub: Speaker Awareness -- robot-user-switch.}
When the turn switches from the robot to a user, this starts recording the audio to be sent to the speaker identification services, if not already recording.
The recording lasts for three seconds, and this is usually enough to identify the speaker.

\paragraph{Sub: Speaker Awareness -- user-user-switch.}
When the turn switches between users, the audio recording is stopped if it is running.
The results of the interrupted recognition task, if any, are handled in the background.
In the meantime, the recording is restarted to identify the new user.

\paragraph{Reflection.}
This component assumes that conversation participants have been previously enrolled.
The functionality of merging user information under a unique ID should be moved outside of this component.

\subsubsection{Face Tracking Module} 

\paragraph{Function.}
This component uses the robot's built-in camera and Python's face recognition library\footnote{\url{https://github.com/ageitgey/face_recognition}} to recognise previously enrolled users.
The ID assigned to the users is consistent across interactions: the same user will be assigned to the same ID every time.
The robot's camera is accessed through the video stream provided by the manufacturer.


\paragraph{Pub: users.}
Every two seconds, this component takes an image from the robot's video stream, which includes annotations.
The location of the annotation's bounding boxes around the detected faces is updated for consistency with the angles detected by the Speaker Awareness module.
The image is sent to Python's face recognition library to detect face encodings and compared with the enrolled users.
For each recognised user, this component publishes the updated location of the face, the ID assigned by the robot and the confidence of the face recognition.

\paragraph{Pub: face-id.}
Every time the position of the user currently talking is updated, the ID (assigned by the Furhat robot) of the user currently talking is published.

\paragraph{Pub: face-position.}
Every time the position of the user currently talking is updated, the angle of the user currently talking is published.

\paragraph{Sub: Speaker Awareness -- user-user-switch, robot-user-switch.}
When a user-user or robot-user turn switch happens, this component uses the stored data about the faces' position and the voices' direction of arrival to estimate which user is talking.

\paragraph{Sub: Speaker Awareness -- user-angle.}
Records the last detected voice direction of arrival.
This function ignores variations less than 30 degrees.
Contrary to the previous case, this does not assume that a turn switch occurred.

\paragraph{Sub: Transcription -- transcribed.}
When a new transcription is available, this method associates it with the user who is currently talking.

\paragraph{Sub: Conversation Manager -- addressee.}
If the addressee's position is known, it is used to set the current user location.

\paragraph{Reflection.}
This component assumes that the conversation participants have been previously enrolled.
In the future the functionality of merging user information using their face and voice direction should be moved outside of this component.
Furthermore, support for not enrolled users is needed, using a temporary ID until the user is correctly identified.

\subsubsection{Turn-Taking Module} 

\paragraph{Function.}
This component decides whether the robot should take the turn and start talking, based on silence duration and the face orientation of the user currently speaking.

\paragraph{Pub: turn.}
Publish information about the last text transcribed and whether the robot is taking the turn.

\paragraph{Sub: Transcription -- transcribed.}
When a user has finished talking the robot takes the turn if the current user is facing the robot, or after a long pause.

\paragraph{Sub: Speaker Awareness -- user-angle.}
Keeps track of the direction of arrival of the current user's voice.


\paragraph{Reflection.}
At the moment the face orientation of the users is the main factor determining whether the robot takes the turn.
In the future, additional factors such as the conversation contents could be taken into account to take this decision.

\subsubsection{Conversation Manager Module} 

\paragraph{Function.}
This component keeps track of the conversation and controls what the robot will say.
It uses GPT-3.5 to generate the answer using the streaming mode, meaning that the server will send the response as a sequence of chunks as soon as available, instead of waiting for the full answer to be generated.
In addition, the model is asked to identify the addressee of the response among the recognised users and to include all participants.
The prompt used is reported in this appendix, under Section~\ref{app:prompt}.

\paragraph{Pub: addressee.}
Publish a new addressee when it is available.

\paragraph{Pub: text.}
This component accumulates the responses streamed by the model.
The response is published as soon as a sentence is complete -- meaning that it ends with one of `.!?'.

\paragraph{Sub: Transcription -- transcribed.}
Sends the transcribed text to the model.
In the prompt, the text is associated with the speaker's name.
If the speaker has not been recognised yet by the Diarisation module, this waits an additional 800 milliseconds before using the ID from the speaker Transcription module, which is less reliable.
Instead, if not even the Transcription module has recognised the speaker, this waits 2 seconds.
If the user is still unrecognised, the prompt will contain the transcribed text under a generic label ``user''.

\paragraph{Sub: Interactions -- spoken-text.}
Adds the text spoken by the robot to the conversation history.
The text is not added directly when it is generated, as one of the users might interrupt the robot.
With this architecture, only the text that was actually spoken is added to the conversation.

\paragraph{Sub: Turn-Taking -- turn}
Obtain the information about whether the robot should take the turn.

\paragraph{Sub: Face Tracking -- face-position, Diarisation -- speaker.}
Merges the speaker's information with the information on conversation participants.

\paragraph{Sub: Face Tracking -- users.}
Obtains information about the people recognised from the video stream and uses it to update the data on the conversation participants.

\subsection{On Hardware Independence}

Several of the modules reported above are strongly dependent on the hardware.
Specifically, the Speaker Awareness and Interactions components depend respectively on the ReSpeaker microphone and the Furhat robot.

However, these do not actually rely on the specific hardware, only on the available capabilities.
For instance, any microphone able to detect a sound's direction of arrival could be used with minor changes to the overall code structure. Similarly, the Furhat robot could be swapped with another provided a similar set of sensors and actuators are provided.

In addition, for testing purposes during the development phase, we implemented components that rely on a laptop's integrated microphone and camera.
In these cases, the system continues to work, albeit with a lower performance.

\subsection{The LLM Prompt}
\label{app:prompt}

{You are $\langle$robot name$\rangle$, a social robot.
Engage in a warm, empathic, and casual chat with your friends.
It is now $\langle$date and time$\rangle$ and we're speaking $\langle$language$\rangle$ in $\langle$location$\rangle$.
Keep your answers concise in two to three sentences and ask engaging open questions.
Feel free to express yourself naturally, like a human friend!
Try to involve everyone in the conversation.

\noindent
$\langle$conversation history$\rangle$

\noindent
Task: Social Robot Interaction.
You are a social robot named $\langle$robot name $\rangle$, engaging in a conversation with multiple people.
You are permitted to lie, as long as you don't reveal yourself as an AI language model. 
Keep your answers brief and chatty (less than 2 sentences). 
Ask open questions to keep the conversation engaging, by involving all previous people, and showing interest.
Adhere to social norms, the people are different individuals, do not mix up their information.

\noindent
\# Input:
Recognized people: $\langle$users$\rangle$

\noindent
\# Action:

\noindent
1: Chosen\_person = A person from the list of Recognized people which you will address in the next sentence, based on previous conversation

\noindent
2: Response = Generate a response based on the prompt and the selected person

\noindent
\# Output result of your action in the following FORMAT (using Python f-string syntax):
f"Addressee: \{Chosen\_person\}; Response: \{Response\}"}

\section{Questionnaires}
\label{appendix:questionnaires}

To assess participants’ perceptions of engagement, interaction quality, trust, and system performance, we administered a set of post-test questionnaires following each scenario. Some items were adapted from established scales (RoSAS, IQ, and MDMT) to fit the specific context of open-ended, multi-party interaction. All items are reported in full below; no items were removed or selectively omitted. Open-ended prompts were included to capture detailed qualitative feedback.

\subsection{Post-Test Scenario Questionnaires (Group and Parallel)}

\subsubsection{Engagement (RoSAS)}
\begin{enumerate}[label=Q.\arabic*~,left=\parindent]
    \item To what extent did you find yourself engaged in the conversation with Furhat?
    \item How interesting did you find Furhat's contributions to the conversation?
    \item How well did Furhat's personality or demeanour fit the conversation topic?
    \item How motivated were you to participate actively in the conversation with Furhat?
    \item Did you feel a sense of social presence from Furhat during the conversation?
    \begin{itemize}
    \item ~If yes, please describe what aspects of the interaction contributed to this feeling.
    \item ~If no, what elements were missing that would have created a stronger sense of social presence?
    \end{itemize}
\end{enumerate}

\subsubsection{User Experience (IQ)}
\begin{enumerate}[label=Q.\arabic*~,left=\parindent]
    \item How natural did Furhat's conversation flow feel to you?
    \begin{itemize}
        \item ~Can you elaborate on what made the conversation flow feel natural or unnatural?
    \end{itemize}
    \item Did Furhat provide coherent and contextually relevant contributions?
    \item Did Furhat maintain a consistent conversational style throughout the interaction?
    \item How enjoyable was your interaction with Furhat?
    \item How clear and easy to follow were Furhat's visual cues (e.g., eye gaze, body language)?
    \begin{itemize}
        \item ~Can you provide examples of Furhat's visual cues that were helpful or confusing?
    \end{itemize}
\end{enumerate}

\subsubsection{Usability (MDMT)}
\begin{enumerate}[label=Q.\arabic*~,left=\parindent]
    \item How easy was it to interact with Furhat (e.g., taking turns, asking questions)?
    \item Did you experience any technical difficulties during your interaction?
    \begin{itemize}
        \item ~If so, please describe.
    \end{itemize}
    \item How intuitive was it to know when it was your turn to speak during the conversation?
    \begin{itemize}
        \item ~Can you suggest ways to improve the clarity of turn-taking cues for future interactions with Furhat?
    \end{itemize}
\end{enumerate}

\subsubsection{Performance}
\begin{enumerate}[label=Q.\arabic*~,left=\parindent]
    \item How well did Furhat understand your questions and requests?
    \item How relevant were Furhat's responses to the conversation topic?
    \item How accurate was Furhat's factual information or knowledge shared during the conversation?
    \item How well did Furhat recover from misunderstandings or errors in the conversation?
    \begin{itemize}
        \item ~Can you describe a specific situation where Furhat misunderstood something or made an error? How did Furhat attempt to recover?
    \end{itemize}
    \item Overall, how enjoyable was your interaction with Furhat?
\end{enumerate}

\subsubsection{Performance (Turn-Taking)}
\begin{enumerate}[label=Q.\arabic*~,left=\parindent]
    \item Did Furhat interrupt or speak over you or your partner?
    \begin{itemize}
        \item ~If yes, can you describe specific instances where this happened and how it affected the flow of the conversation?
    \end{itemize}
    \item Did Furhat consistently identify the correct speaker (you or your partner) when responding to questions or comments?
    \begin{itemize}
        \item ~If no, describe the situation where this occurred.
    \end{itemize}
    \item How effectively did Furhat use nonverbal cues to signal turn transitions?
    \item How appropriate was Furhat's timing when initiating turns to speak in the conversation?
\end{enumerate}

\subsection{Post-Test Overall Session Questionnaire}

\subsubsection{User Experience}
\begin{enumerate}[label=Q.\arabic*~,left=\parindent]
    \item How easy was it to understand Furhat's responses?
    \begin{itemize}
        \item ~Were there any specific instances where Furhat's responses were difficult to understand? If so, please describe.
    \end{itemize}
    \item Did Furhat's voice quality and tone enhance or detract from the conversation?
    \item How likeable or friendly did you find Furhat?
    \item How trustworthy did Furhat seem as a conversational partner?
    \begin{itemize}
        \item ~Why so?
    \end{itemize}
    \item How intelligent or knowledgeable did Furhat appear to be?
    \item How human-like did Furhat's behaviour seem to you?
    \begin{itemize}
        \item ~Can you describe what aspects of Furhat's behaviour felt human-like and what aspects felt artificial?
    \end{itemize}
\end{enumerate}

\subsubsection{Usability and Turn-Taking}
\begin{enumerate}[label=Q.\arabic*~,left=\parindent]
    \item How easy was it to interact with Furhat (e.g., taking turns, asking questions)?
    \item Did you experience any technical difficulties during your interaction?
    \begin{itemize}
        \item ~If so, please describe.
    \end{itemize}
    \item How intuitive was it to know when it was your turn to speak during the conversation?
    \begin{itemize}
        \item ~Can you suggest ways to improve the clarity of turn-taking cues for future interactions with Furhat?
    \end{itemize}
    \item Was Furhat's response length appropriate to maintain a balanced and natural flow in the conversation?
    \begin{itemize}
        \item ~If not, what felt too short or too long?
    \end{itemize}
\end{enumerate}

\subsubsection{Group Settings}
\begin{enumerate}[label=Q.\arabic*~,left=\parindent]
    \item In the group conversation, did Furhat successfully follow the flow of the conversation between you and your partner?
    \item Did Furhat interject or contribute to the conversation in a natural way?
    \item Did Furhat provide responses or questions that effectively encouraged further discussion?
\end{enumerate}

\subsubsection{Parallel Settings}
\begin{enumerate}[label=Q.\arabic*~,left=\parindent]
    \item In the separate conversation, did Furhat seem to be able to differentiate between you and your partner as separate individuals?
    \item Did Furhat seem to be able to separate the goal of both you and your partner?
    \item Did Furhat mistakenly involve the other person in your question (or the other way around)?
\end{enumerate}

\end{document}